\documentclass[12pt]{iopart}

\usepackage{graphicx}
\usepackage{here}
\usepackage{stackrel}
\usepackage{amssymb,accents,latexsym}
\usepackage{braket}

\usepackage{subfigure}
\usepackage{tikz}

\newcommand{\QCD}{{\textrm{\tiny QCD}}}

\newcommand{\BG}{{\textrm{\tiny BG}}}
\newcommand{\TF}{{\textrm{\tiny TF}}}
\begin{document}

\title[Fractal structure of Yang-Mills fields]{Fractal structure of Yang-Mills fields}

\author{Airton~Deppman$^{1}$, Eugenio~Meg\'{\i}as$^{2}$ and D\'ebora P.~Menezes$^{3}$}
\address{
$^{1}$ Instituto de F\'{\i}sica, Rua do Mat\~ao 1371-Butant\~a, S\~ao Paulo-SP, CEP 05580-090, Brazil \\
$^{2}$ Departamento de F\'{\i}sica At\'omica, Molecular y Nuclear and Instituto Carlos I de F\'{\i}sica Te\'orica y Computacional, Universidad de Granada, \\
Avenida de Fuente Nueva s/n, 18071 Granada, Spain \\
$^{3}$ Departamento de F\'{\i}sica, CFM-Universidade Federal de Santa Catarina, Florian\'opolis, SC-CP. 476-CEP 88.040-900, Brazil
}
\ead{deppman@usp.br, emegias@ugr.es,  debora.p.m@ufsc.br}

\vspace{10pt}
%\begin{indented}
%\item[]August 2017
%\end{indented}

\begin{abstract}
The origin of non-extensive thermodynamics in physical systems has
been under intense debate for the last decades. Recent results
indicate a connection between non-extensive statistics and
thermofractals. After reviewing this connection, we analyze how
scaling properties of Yang-Mills theory allow the appearance of
self-similar structures in gauge fields. The presence of such
structures, which actually behave as fractals, allows for recurrent
non-perturbative calculations of vertices. It is argued that when a
statistical approach is used, the non-extensive statistics is
obtained, and the Tsallis entropic index, $q$, is deduced in terms of
the field theory parameters. The results are applied to QCD in the
one-loop approximation, resulting in a good agreement with the value
of $q$ obtained experimentally.
\end{abstract}

\noindent{\it Keywords}: Yang-Mills theory, non-extensive statistics, thermofractals, quark-gluon plasma, hadron physics
%\maketitle

\section{Introduction}
\label{sec:introduction}

Fractals are complex systems with internal structure presenting scale invariance and self-similarity~\cite{Mandelbrot}. One of the most important features of fractals are the scaling properties, where the internal structure of the fractal is equal to the main fractal but with a reduced scale. Tsallis statistics was introduced as a generalization of Boltzmann-Gibbs (BG) statistics~\cite{Tsallis:1987eu} by considering a non-additive form of entropy, and it has found wide applicability in the last few years, see e.g. Refs.~\cite{Tempesta:2011vc,Kalogeropoulos:2014mka}. However, the full understanding of this statistics has not been accomplished yet. 

On the other hand, Yang-Mills field theory is a prototype theory that allows the description of three among the four known fundamental interactions. Renormalization group invariance is a fundamental aspect of the Yang-Mills theory, playing an important role in the renormalization of the theory after the ultraviolet (UV) divergences are subtracted~\cite{Dyson:1949ha,GellMann:1954fq}.

The goal of this manuscript is to show the link between fractals, Tsallis non-extensive statistics, and  renormalization group invariance of Yang-Mills theory. We will analyze how scaling properties of a Yang-Mills theory lead to recurrence relations, which amount to a self-similar behavior of $n$-point diagrams by scale evolution.

\section{Tsallis statistics in high energy physics}
\label{sec:Tsallis_highenergy}

\begin{figure*}[t]
\centering
 \begin{tabular}{c@{\hspace{3.5em}}c}
 \includegraphics[width=0.46\textwidth]{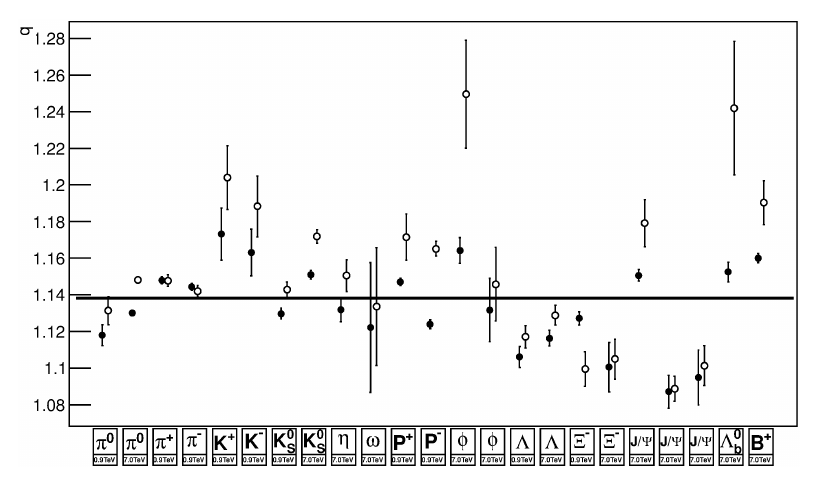} &
\includegraphics[width=0.42\textwidth]{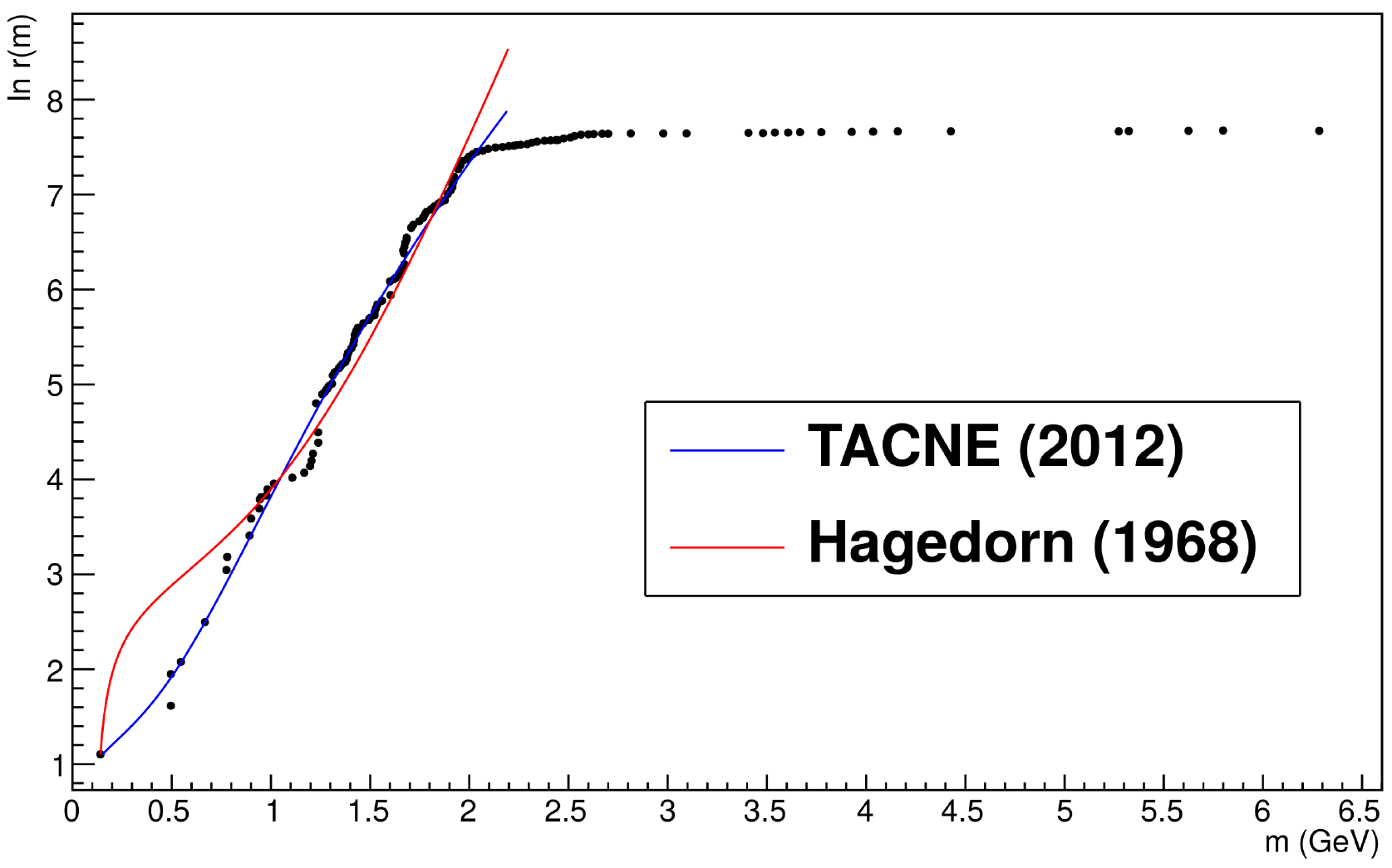}
\end{tabular}
 \caption{\it Left panel: Entropic index, $q$, obtained from a fit of the abundance of different hadron species in $pp$ collisions using Tsallis statistics, cf. Refs.~\cite{Marques:2012px,Marques:2015mwa}. Right panel: Cumulative hadron spectrum, as a function of the hadron mass. The dots stand for the PDG result. We display also he result by using the non-extensive self-consistent thermodynamics (blue) and the result predicted by Hagedorn (red), cf. Ref.~\cite{Marques:2012px}.}
\label{fig:experiment}
\end{figure*}

R. Hagedorn proposed a self-consistent thermodynamical approach to high energy collisions based on the BG statistics~\cite{Hagedorn:1965st} that predicts an exponential behavior of the statistical distributions in energy and momentum of particle production of hadron species in high energy $pp$ collisions. This approach, however, disagrees from experimental data as these tend to behave instead as a power-law. The extension of the Hagedorn theory to non-extensive statistics was studied in Ref.~\cite{Deppman:2012us}, based on the use of the Tsallis factor
\begin{equation}
P(\varepsilon) = A \left[ 1+(q-1)\frac{\varepsilon}{k\tau} \right]^{-\frac{1}{q-1}} \,,
\end{equation}
instead of the exponential BG factor, which allowed to reproduce the distribution of all the species produced in $pp$ collisions with a high accuracy, leading to the result~\cite{Marques:2012px,Marques:2015mwa} 
\begin{equation}
q = 1.14 \pm 0.01 \,,
\end{equation}
(see left panel of Fig.~\ref{fig:experiment}). Moreover, this extension predicted also a power-law behavior for the hadron spectrum, i.e.
\begin{equation}
\rho(m)= \rho_o \left[1+(q-1) \frac{m}{M} \right]^{\frac{1}{q-1}} \,.
\end{equation}
The comparison of this distribution with the hadron spectrum from the Particle Data Group (PDG), as displayed in the right panel of Fig.~\ref{fig:experiment}, leads to an important improvement with respect to the exponential distribution $\rho(m) = \rho_o \, e^{M/T_H}$ proposed by Hagedorn, specially at the lowest masses, cf. Ref.~\cite{Marques:2012px}.

Apart from these considerations, there are many other applications in which Tsallis statistics play an important role. This includes high energy collisions~\cite{Cleymans:2011in,Marques:2015mwa,Wong:2015mba}, hadron models~\cite{Cardoso:2017pbu}, hadron mass spectrum~\cite{Marques:2012px}, neutron stars~\cite{Menezes:2014wqa}, lattice QCD~\cite{Deppman:2012qt}, non-extensive statistics~\cite{Deppman:2012us,Megias:2015fra,Deppman:2017fkq}, and many others. In the next section we will see that the power-law behavior of Tsallis statistics leads, in fact, to a subtle connection with thermofractals.

\section{Tsallis statistics and thermofractals}
\label{sec:Tsallis_Statistics}

In this section we will provide a short introduction to the formalisms of Tsallis statistics and thermofractals, and discuss the link between both descriptions.

\subsection{Tsallis statistics}
\label{subsec:Tsallis}

Tsallis statistics is a generalization of BG statistics, under the assumption that the entropy of the system is non-additive. For two independent systems $A$ and $B$
\begin{equation}
S_{A+B} = S_A + S_B + (1-q)S_A S_B \,,
\end{equation}
where the entropic index, $q$, measures the degree of non-extensivity~\cite{Tsallis:1987eu}. If we define the $q$-exponential and $q$-logarithmic functions as
\begin{equation}
e_q^{(\pm)}(x)=[1 \pm (q-1)x]^{\pm1/(q-1)}\,, \quad \log^{(\pm)}_q(x)=\pm (x^{\pm(q-1)}-1)/(q-1) \,,
\end{equation}
where $(+)$ in $e_q(x) \, (\log_q(x))$ stands for $x \ge 0 \, (1)$, and $(-)$ stands for $x < 0 \, (1)$, then the grand-canonical partition function for a non-extensive ideal quantum gas is given by~\cite{Megias:2015fra}
\begin{equation}
 \log\Xi_q(V,T,\mu) =
 -\xi V\int \frac{d^3p}{{(2\pi)^3}} \sum_{r=\pm}\Theta(r x)\log^{(-r)}_q \left( \frac{ e_q^{(r)}(x)-\xi}{ e_q^{(r)}(x)} \right)  \,,
\end{equation}
where $x = (E_p - \mu)/(kT)$, the particle energy is $E_p = \sqrt{p^2+m^2}$, with $m$ being the mass and $\mu$ the chemical potential, $\xi = \pm 1$ for bosons and fermions respectively, and $\Theta$ is the step function. Note that $e_q^{(\pm)}(x) \stackrel[q \to 1]{\longrightarrow}{} e^x \quad $ and $\quad \log_q^{(\pm)}(x) \stackrel[q \to 1]{\longrightarrow}{} \log(x)$, so that Tsallis statistics reduces to BG statistics in the limit $q \to 1$. This formalism has been used to successfully describe the thermodynamics of Quantum Chromodynamics (QCD) in the confined phase by using the hadron resonance gas approach, with applications in high energy physics~\cite{Megias:2015fra}, hadron physics~\cite{Andrade:2019dgy}, and neutron stars~\cite{Menezes:2014wqa}.

\subsection{Thermofractals}
\label{subsec:Thermofractals}

The emergence of the non-extensive behavior has been attributed to
different causes. These include long-range interactions, correlations and memory
effects~\cite{Borland:1998}, temperature fluctuations, and finite size of the
system, among others. We will show in this section that   a natural derivation of non-extensive statistics in terms of thermofractals is possible. These are systems in thermodynamical equilibrium presenting the following properties~\cite{Deppman:2016fxs}:
\begin{enumerate}
\item Total energy is given by~$U = F + E$, where $F$ is the kinetic energy, and $E$ is the internal energy of $N$ constituent subsystems, so that $E = \sum_{i=1}^N \varepsilon_i^{(1)}$.
\item The constituent particles are thermofractals. This means that the energy distribution $P_{\TF}(E)$ is self-similar or self-affine, i.e. at level $n$ of the hierarchy of subsystems, $P_{\TF (n)}(E)$ is equal to the distribution in the other levels.
\end{enumerate}
The energy distribution according to BG statistics is given by
\begin{equation}
 P_{\BG}(U) dU=A \exp(-U/kT) dU \,, \label{eq:P_BG}
\end{equation}
where $A$ is a normalization constant. In the case of thermofractals the phase space must include the momentum degrees of freedom $(\propto f(F))$ of free particles as well as internal degrees of freedom $(\propto f(\varepsilon))$. Then, the internal energy is $dE \propto [P_{\TF}(\varepsilon)]^\kappa d\varepsilon$ where $\kappa$ is an exponent to be determined, and one has~\cite{Deppman:2016fxs}
\begin{equation}
  P_{\TF (0)}(U) dU = A^\prime  F^{\frac{3N}{2}-1} \exp\left( -\frac{\alpha F}{kT} \right) dF \left[ P_{\TF (1)}(\varepsilon) \right]^\kappa d\varepsilon \,,  \label{eq:PTF01_1}
\end{equation} 
with $\alpha = 1 + \varepsilon/(kT)$ and $\varepsilon/(kT) = E/F$. This expression relates the distributions at level $0$ and $1$ of the subsystem hierarchy. After integration, and by imposing self-similarity, i.e.
\begin{equation}
P_{\TF (0)}(U) \propto P_{\TF (1)}(\varepsilon) \,, \label{eq:PTF01_2}
\end{equation}
 the simultaneous solution of Eqs.~(\ref{eq:PTF01_1}) and (\ref{eq:PTF01_2}) is obtained with~\cite{Deppman:2017fkq}
\begin{equation}
 P_{\TF (n)}(\varepsilon)= A_{(n)} \cdot \left[1 + (q-1) \frac{\varepsilon}{k\tau} \right]^{-\frac{1}{q-1}} \,.  \label{eq:selfsimilar}
\end{equation}
We find that the distribution of thermofractals obeys Tsallis statistics with $q-1 = 2(1-\kappa)/(3N)$ and $\tau = N (q-1) T$. %This result shows that there is a clear connection between hadron structure, which is described by Tsallis statistics, and thermofractals. 
In the rest of this manuscript, we will study the connection between thermofractals and quantum field theory, and in particular to QCD.

\section{Scales in Yang-Mills theory}
\label{sec:scales_YM}

As discussed in Secs.~\ref{sec:Tsallis_highenergy} and \ref{sec:Tsallis_Statistics} the phenomenology of QCD, and in particular its thermodynamics,  can be described by Tsallis statistics. In addition, we have seen in Sec.~\ref{sec:Tsallis_Statistics} that thermofractals obey Tsallis statistics. Then, a natural question arises: Is it possible a thermofractal description of Yang-Mills theory? In Secs.~\ref{sec:scales_YM} and \ref{sec:Fractal_YM} we will try to address this interesting issue. We will start by studying the scaling properties in Yang-Mills theory.

\subsection{Renormalization of gauge fields}
\label{subsec:renormalization}

The simplest non-Abelian gauge field theory has Lagrangian density including bosons and fermions given by 
\begin{equation}
{\cal L}=-\frac{1}{4} F^{a}_{\mu \nu} F^{a \mu \nu}+ i \bar{\psi_j} \gamma_{\mu} D^{\mu}_{ij} \psi_j \,,
\end{equation}
where~$F^{a}_{\mu \nu}=\partial_{\mu} A^{a}_{\nu} - \partial_{\nu} A^{a}_{\mu}+g f^{abc}A^{b}_{\mu}A^{c}_{\nu}$ is the field strength of the gauge field, and~$D^{\mu}_{ij}= \partial_{\mu}\delta_{ij}-igA^{a \, \mu} T^{a}_{ij}$ is the covariant derivative,  with $f^{abc}$ being the structure constants of the group, and $T^{a}$ the matrices of the group generators. $\psi$ and $A$ represent, respectively, the fermion and the vector fields.

\begin{figure*}[t]
\centering
 \begin{tabular}{c@{\hspace{3.5em}}c}
 \includegraphics[width=0.37\textwidth]{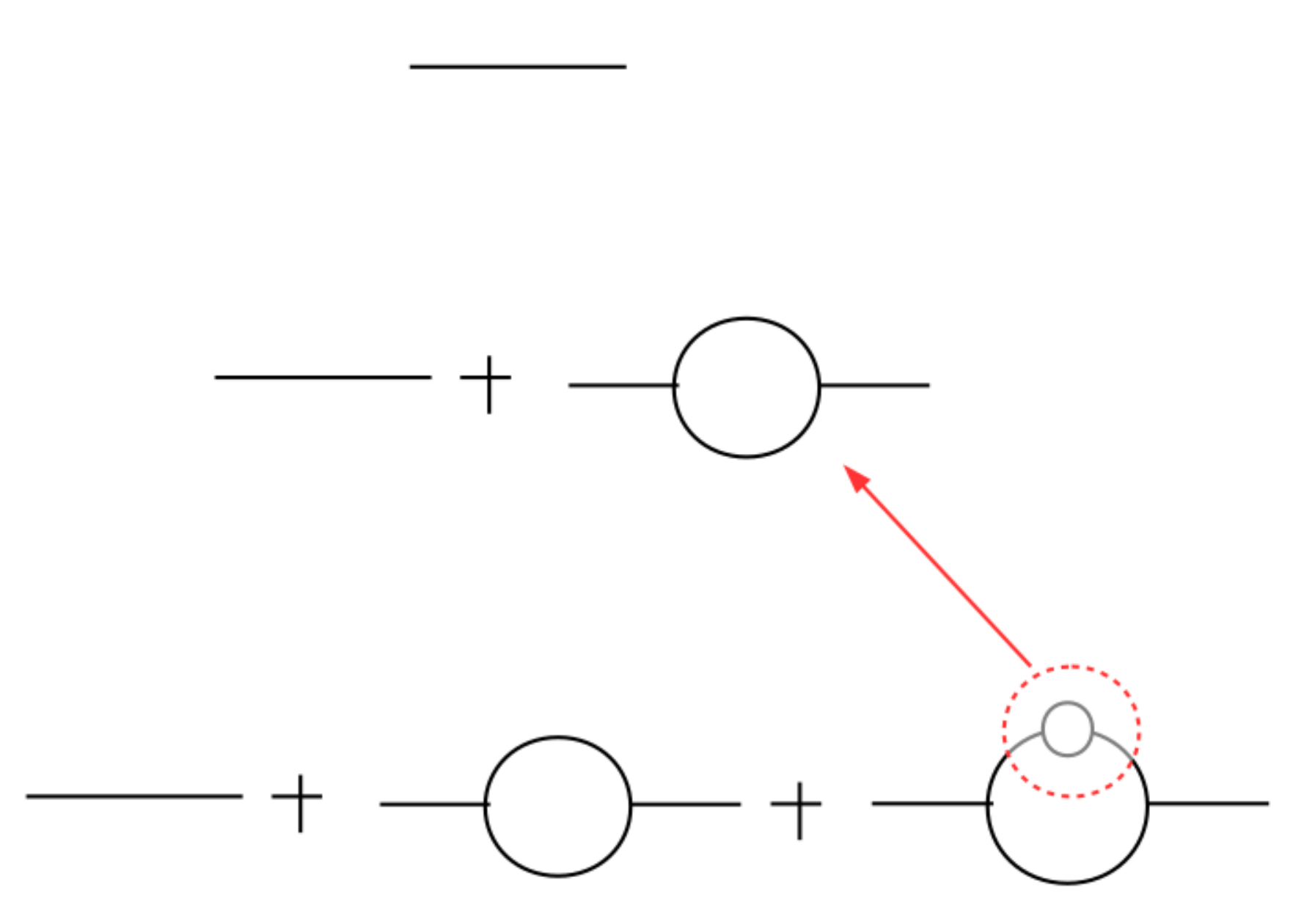} &
\includegraphics[width=0.26\textwidth]{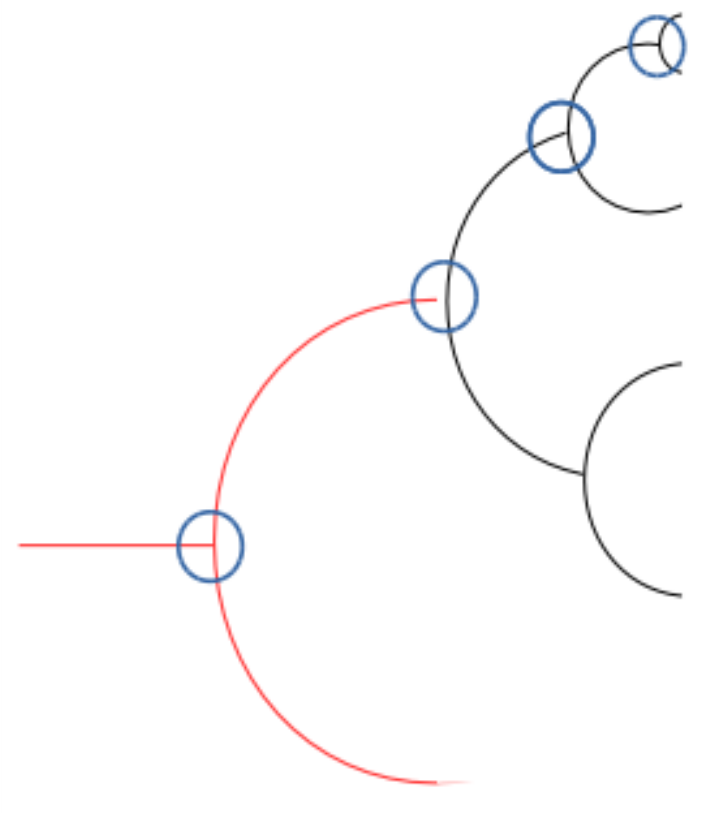} 
\end{tabular}
%\vspace{-4cm}
 \caption{\it Left panel: Diagrams showing the scaling properties of Yang-Mills fields. It is shown the loops at different orders in the perturbative expansion. Right panel: Diagrams of multiparticle production in $pp$ collisions. The initial parton (first black line from the left) may be considered as a constituent of another parton (first red line from the left). 
 The effective vertex couplings, which are given by the second expression in Eq.~(\ref{eq:effective_coupling}) are indicated by circles.}
\label{fig:scaling_multiparticle}
\end{figure*}

This theory is renormalizable, which means that the UV regularized vertex functions are related to the renormalized vertex functions with renormalized parameters, $\bar{m}$ and $\bar{g}$, as~\cite{Dyson:1949ha,GellMann:1954fq}
\begin{equation}
\Gamma(p,m,g)=\lambda^{-D} \Gamma(p,\mu,\bar{g}) \,.
\end{equation}
This property is described by the renormalization group equation, also known as Callan-Symanzik (CS) equation, which is given by~\cite{Callan:1970yg,Symanzik:1970rt}:
\begin{equation}
\left[M\frac{\partial}{\partial M}  + \beta_{\bar g} \frac{\partial}{\partial \bar{g}}  + \gamma \right]\Gamma=0 \,,
\end{equation}
where $M$ is the scale parameter, and the beta function is defined as $\beta_{\bar g} = M \frac{\partial \bar g}{\partial M}$. As it is shown in the left panel of Fig.~\ref{fig:scaling_multiparticle}, renormalization group invariance means that, after proper scaling, the loop in a higher order graph in the perturbative expansion is identical to a loop in lower orders. This is a direct consequence of the CS equation, and it is of fundamental importance in what follows. 

\subsection{Multiparticle production}
\label{subsec:multiparticle}

An interesting physical situation in which these scaling properties emerge is in multiparticle production in $pp$ collisions~\cite{Konishi:1978ks}. We show in the right panel of Fig.~\ref{fig:scaling_multiparticle} a typical diagram in which two partons are created in each vertex. When effective masses and charges are used, the line of the respective field in Feynman diagrams represents an effective particle. An irreducible graph represents an effective parton, and the vertices are related to the creation of an effective parton. Due to the complexity of the system, it is desirable a statistical description in which the summation of all the diagrams becomes equivalent to an ideal gas of particles with different masses~\cite{Dashen:1969ep,Venugopalan:1992hy}. Another example is given by the hadron structure: as in the case of multiparticle production, too many complex graphs should be considered when studying the hadron at a high resolution. This is why calculations are typically limited either to the first leading orders, or to lattice QCD methods.

\section{Fractal structure of gauge fields}
\label{sec:Fractal_YM}

We will introduce in this section the formalism that will allow us to understand the fractal structure of gauge theories.

\subsection{Statistical description of the partonic state}
\label{subsec:partonic} 

We will consider that at any scale the system can be described as an ideal gas of particles with different masses, i.e masses might change with the scale. The time evolution of an initial partonic state is given by
\begin{equation}
\ket{\Psi} \equiv \ket{\Psi(t)}=e^{-iHt} \ket{\Psi_o} \,.
\end{equation}
The state $\ket{\Psi}$ can be written as  $\ket{\Psi}=\sum_{\{n\}} \braket{\Psi_n|\Psi} \ket{\Psi_n}$, where $\ket{\Psi_n}$ is a state with $n$ interactions in the vertex function. Each proper vertex gives rise to a term in the Dyson series, and at time~$t$ the partonic state is given by
\begin{equation}
 \ket{\Psi} =  \sum_{\{n\}} (-i)^n \int dt_1 \dots dt_n  e^{-iH_o(t_n-t_{n-1})} g  \dots e^{-iH_o (t_1-t_o)}  \ket{\Psi_o} \,,
\end{equation} 
where $g$ represents the interaction and $t_n>t_{n-1}>\dots>t_1 >t_o$, while $\sum_{\{n\}}$ runs over all possible terms with $n$ interaction vertices. Let us introduce states of well-defined number of effective partons, $\ket{\psi_N}$, so that
\begin{equation}
 \ket{\Psi_n}=\sum_{N} \braket{\psi_N|\Psi_n} \ket{\psi_N}\,. 
\end{equation}
Therefore $\ket{\psi_N}={\cal S} \ket{\gamma_1,m_1,p_1, \dots , \gamma_N,m_N,p_N}$, where $m_i$ and $p_i$ are the mass and momentum of the $i$ partonic state, and $\gamma_i$ represents all relevant quantum numbers necessary to completely characterize the partonic state. These states can be understood as a quantum gas of particles with different masses. Let us remark at this point that the number of particles in the state $\ket{\Psi_n}$ is not directly related to $n$, since high order contributions to the $N$ particles states can be important. The rule is $N \le N_{\max}(n):=n(\tilde{N}-1)+1$, where $\tilde{N}$ is the number of particles created or annihilated at each interaction. In Yang-Mills field theory, $\tilde{N}=2$.

Let us study the probability to find a state with one parton with mass between $m_o$ and $m_o + dm_o$, and momentum between $p_o$ and $p_o + dp_o$. This is given by
\begin{equation}
\braket{\gamma_o,m_o,p_o, \dots|\Psi(t)} = \sum_n  \sum_{N} \braket{\Psi_n|\Psi(t)} \braket{\psi_N|\Psi_n} \braket{\gamma_o,m_o,p_o,\dots|\psi_N} \,. \label{eq:probPsi}
\end{equation}
There are three factors in the rhs of this equation. The first one, $\braket{\Psi_n|\Psi(t)}$, is related to the probability that an effective parton with energy between $E$ and $E + dE$ at time $t=0$ will evolve in such a way that at time $t$ it will generate an arbitrary number of secondary effective partons in a process with $n$ interactions. This factor can be written as $\braket{\Psi_n|\Psi(t)}=G^n P(E) dE$, where $P(E)$ is the probability distribution of the initial particle, and $G^n$ is the probability that exactly $n$ interactions will occur in the elapsed time. The second bracket is the probability to get the configuration with $N$ particles after $n$ interactions, i.e.
\begin{equation}
\braket{\psi_N|\Psi_n}=C_N(n) \stackrel[n \gg 1]{\simeq}{}   \left( \frac{N}{n(\tilde{N}-1)} \right)^4 \,.
\end{equation}
Finally, the last bracket in Eq.~(\ref{eq:probPsi}) can be calculated statistically, leading to the following result~\cite{Deppman:2019yno}
\begin{equation}
\braket{\gamma_o,m_o,p_o,\dots|\psi_N} \simeq A(N) P_N\left( \frac{\varepsilon_j}{E} \right)  d^4\left( \frac{p_j}{E} \right) \,,
\end{equation}
with
\begin{equation}
A(N) = \frac{\Gamma(4N)}{8\pi \Gamma(4(N-1))}  \qquad \textrm{and} \qquad P_N\left( x \right) = (1-x)^{4N-5} \,,
\end{equation}
where $\Gamma(x)$ is the Euler gamma function, $p_j^\mu = (p_j^0, \vec{p}_j)$ is the fourth momentum of particle $j$ inside the system of $N$~particles, and $\varepsilon_j = p_j^0$ is the energy of that particle. Note that we are not assuming a fixed value for the mass $m_j$ of particle $j$, where $m_j^2 = p^\mu p_\mu$, so that $p_j^0$ and $\vec{p}_j$ are variables that may change independently each other. By combining all these results in Eq.~(\ref{eq:probPsi}), one finally obtains~\footnote{We have used in Eq.~(\ref{eq:probPsi_result}) that for $N$ sufficiently large and $x \ll 1$, one can approximate $\left( 1 - x  \right)^{(4N-5)}  \simeq \left( 1 + x  \right)^{-(4N-5)}$.}
\begin{eqnarray}
\hspace{-2.0cm}  \tilde{P}(\varepsilon)
d^4p_o dE &\equiv& \braket{\gamma_o,m_o, \dots|\Psi(t)} = \nonumber \\ 
\hspace{-2.0cm} &&= \sum_n \sum_{N} G^n
\left(\frac{N}{n(\tilde{N}-1)}\right)^4
\left(1+\frac{\varepsilon}{E}\right)^{-(4N-5)} d^4\left(\frac{p
}{E}\right) P(E)dE \,. \label{eq:probPsi_result}
\end{eqnarray}
As we will show below, the distributions $\tilde{P}(\varepsilon)$ and $P(E)$ can be obtained from considerations about self-similarity.

\subsection{Self-similarity and fractal structure}
\label{subsec:self_similarity}

The energy distribution of a parton, as given by Eq.~(\ref{eq:probPsi_result}), depends on the ratio $\chi = \varepsilon_j / E$. Let us consider that the system with energy $E$ in which the parton with energy~$\varepsilon_j$ is one among $N$ constituents, is itself a parton inside a larger system with energy $\cal{M}$. Then self-similarity implies that
\begin{equation}
\tilde{P}\left( \frac{\varepsilon_j}{E} \right) \propto P\left( \frac{E}{\cal{M}} \right) \,.
\end{equation}
After some algebra, it can be shown that
\begin{equation}
P\left( \frac{\varepsilon}{\lambda}\right) = \left[ 1 + (q-1) \frac{\varepsilon}{\lambda} \right]^{-\frac{1}{q-1}} \,, \label{eq:Pe}
\end{equation}
where $q-1 = (1-\nu)/(4N-5)$, while $\nu$ represents the fraction of total number of d.o.f. of the state $\ket{\psi_N}$ that is involved in each interaction, and $\lambda = (q-1)\Lambda$ is a reduced scale. This probability distribution is a power-law function, and it corresponds exactly to the distribution derived in Eq.~(\ref{eq:selfsimilar}) for thermofractals, cf. Refs.~\cite{Deppman:2016fxs,Deppman:2017fkq}.

Then, an interesting interpretation of the entropic index, $q$, arises: it is related to the number of internal degrees of freedom in the fractal structure, and the distribution of Eq.~(\ref{eq:Pe}) describes how the energy received by the initial parton flows to its internal d.o.f.. In the context of the theory developed in this section, this probability describes how the energy flows from the initial parton to partons at higher perturbative orders. Since new orders are associated to new vertices, this suggests that this distribution plays the role of an effective coupling constant in the vertex function, i.e.
\begin{equation}
 \Gamma=\braket{\Psi_{n+1}| g e^{iH_ot_{n+1}}|\Psi_n} \quad \textrm{with} \quad g=\prod_{i=1}^{\tilde{N}}G\left[1+(q-1)\frac{\varepsilon_i}{\lambda}\right]^{-\frac{1}{q-1}}\,. \label{eq:effective_coupling}
\end{equation}
The situation is schematized in the right panel of Fig.~\ref{fig:scaling_multiparticle}, where the vertex functions that are responsible for the scaling properties of the theory in multiparticle production (as discussed in Sec.~\ref{sec:scales_YM}) are shown.

\section{Effective coupling and beta function}
\label{sec:beta_function}

The CS equation together with the renormalized vertex functions were used to derive the beta function of QCD, which allowed to show that QCD is asymptotically free~\cite{Politzer:1974fr,Gross:1974cs}. The result at the one-loop approximation is
\begin{equation}
\beta_{\QCD} = - \frac{\bar{g}^3}{16\pi^2} \left[ \frac{11}{3} c_1 - \frac{4}{3}c_2 \right] \,, \label{eq:betaQCD}
\end{equation}
where $c_1 \delta_{ab} = f_{acd} f_{bcd}$ and $c_2 \delta_{ab} = \Tr\left( T_a T_b\right)$. Quantitatively, the parameters~$c_1$ and $c_2$ are related to the number of colors and flavors by $c_1 = N_c$ and $c_2 = N_f/2$. In this section, we will study the beta function derived with our ansatz, and compare with that in QCD. 

\begin{figure*}[t]
\centering
 \begin{tabular}{c@{\hspace{3.5em}}c}
 \includegraphics[width=0.43\textwidth]{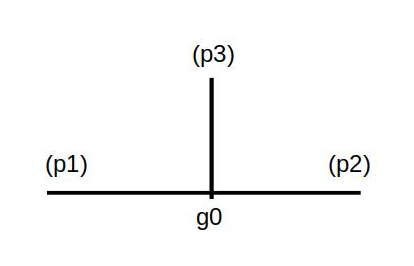} &
\includegraphics[width=0.43\textwidth]{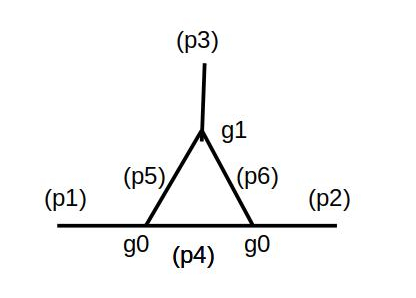} 
\end{tabular}
 \caption{\it Vertex functions at scale $\lambda_o$ (left) and $\lambda$ (right).}
\label{fig:vertex_functions}
\end{figure*}

\begin{figure*}[t]
\centering
 \begin{tabular}{c@{\hspace{3.5em}}c}
 \includegraphics[width=0.33\textwidth]{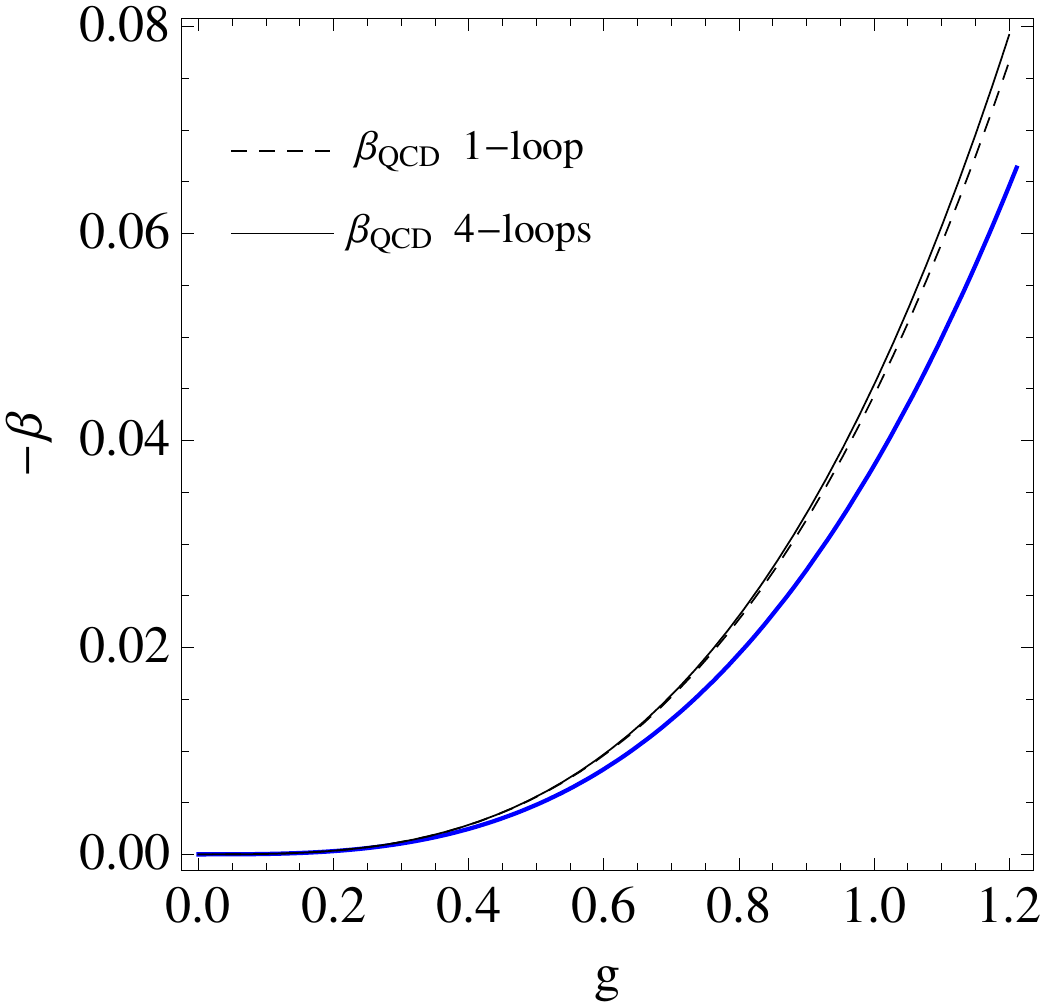} &
\includegraphics[width=0.33\textwidth]{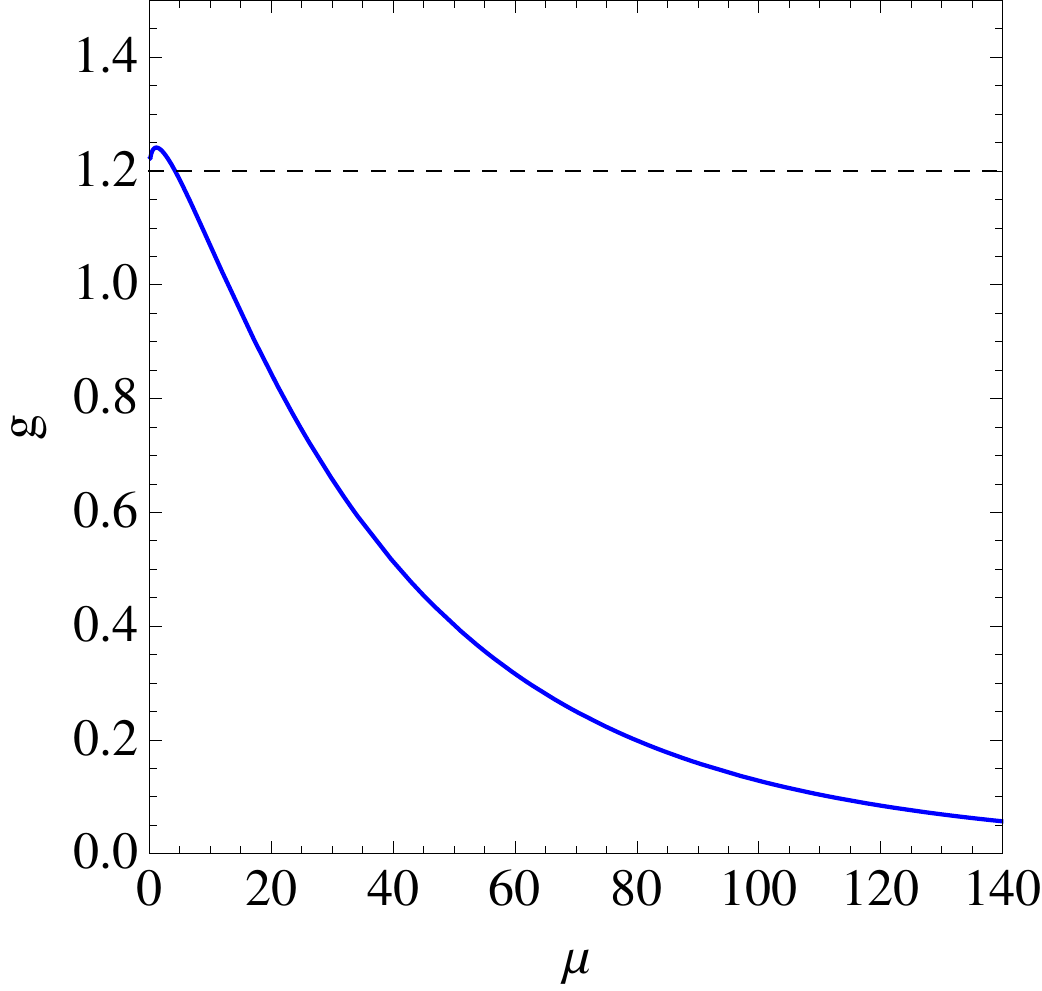}
\end{tabular}
 \caption{\it Left panel: Behavior of beta function against effective  coupling, as given by Eq.~(\ref{eq:betag}). We display also the result from QCD in one-loop~\cite{Politzer:1974fr} and four-loop~\cite{Czakon:2004bu} approximation. Right panel: Effective coupling as a function of the scale~$\mu$, as calculated by Eq.~(\ref{eq:QCDrunningcoupling}).}
\label{fig:betag}
\end{figure*}

Let us consider a vertex in two different orders, as depicted in Fig.~\ref{fig:vertex_functions}. The vertex function at first order, i.e. at scale $\lambda_o$, is
\begin{equation}
 \Gamma_o=\braket{\gamma_2 p_2 , \gamma_3 p_3|g(\lambda_o)e^{iH_ot}| \gamma_1 p_1}\,. \label{eq:initialvertex}
\end{equation}
The next order in the perturbative approximation is given by the vertex with one additional loop at scale~$\lambda$, which results in a vertex function
\begin{eqnarray}
\hspace{-1.5cm} \Gamma &=& \bra{\gamma_2 p_2 , \gamma_3 p_3} g(\lambda_o) e^{iH_ot_3}\ket{\gamma_2 p_6,\gamma_3 p_3, \gamma_4 p_4} \times  \nonumber \\ 
\hspace{-1.5cm} &&\times  \bra{\gamma_2 p_6, \gamma_3 p_3, \gamma_4 p_4} g(\lambda)e^{iH_ot_2}  \ket{\gamma_1 p_5, \gamma_4 p_4} \bra{\gamma_1 p_5, \gamma_4 p_4}g(\lambda_o)e^{iH_ot_1} \ket{ \gamma_1 p_1}\,. \label{eq:expanded1loopvertexfunction}
\end{eqnarray}
By comparing this expression with $\Gamma=\braket{\gamma_2 p_2 , \gamma_3 p_3|\bar{g}\, e^{iH_ot}| \gamma_1 p_1}$, one can identify the effective coupling $\bar{g}$ as
\begin{equation}
 \bar{g}=g(\lambda_o) e^{iH_ot_3}\ket{\gamma_2 p_6,\gamma_3 p_3, \gamma_4 p_4} \Gamma_M \bra{\gamma_1 p_5, \gamma_4 p_4}g(\lambda_o)\,, \label{eq:1loopcoupling}
\end{equation}
where
\begin{equation}
 \Gamma_M=\braket{\gamma_2 p_6, \gamma_3 p_3, \gamma_4 p_4|g(\lambda)e^{iH_ot_2} | \gamma_1 p_5, \gamma_4 p_4}\,. \label{eq:scaledvertex}
\end{equation}
The scaling properties of Yang-Mills fields allow us to relate $\Gamma_M$ to $\Gamma_o$ by an appropriate scale,~$\lambda$. From dimensional analysis, the scaling behavior turns out to be $\Gamma_M(\lambda) = (\lambda/\lambda_o)^4$. Using these considerations and from the CS equation, it results that
\begin{equation}
\beta_{\bar{g}}\frac{\partial \Gamma}{\partial g}=-(d+\gamma_5+\gamma_6)\Gamma\,,
\end{equation}
where $d=4$ and $\gamma_{5,6}$ are the anomalous dimensions. In order to compare with the QCD results, we will study the behavior of $g(\lambda)$ at $\lambda = \lambda_o/\mu$, where $\mu$ is a scaling factor. From Eq.~(\ref{eq:effective_coupling}), one has
\begin{equation}
 g(\mu)=\prod_{i=5}^{6}G\left[1+(q-1)\frac{\varepsilon_i \mu}{\lambda_o}\right]^{-\frac{1}{q-1}}\,, \label{eq:QCDrunningcoupling}
\end{equation}
and substituting it into Eqs.~(\ref{eq:1loopcoupling}) and (\ref{eq:scaledvertex}) one can calculate the beta function in the one loop approximation. By considering the asymptotic limit $(q-1)\mu \gg \lambda_o / \varepsilon_i$, one obtains
\begin{equation}
\beta_{\bar{g}}=-\frac{1}{16\pi^2} \frac{1}{q-1} \bar{g}^{\tilde{N}+1}\,, \label{eq:betag}
\end{equation}
with $\tilde{N} =2$. Finally, by comparing with the QCD result, Eq.~(\ref{eq:betaQCD}), one can make the identification
\begin{equation}
\frac{1}{q-1} = \frac{11}{3}c_1 - \frac{4}{3}c_2 = 7 \,,
\end{equation}
where in the last equality we have used $N_c = N_f/2 = 3$. This result leads to $q = 1.14$, which shows an excellent agreement with the experimental data analyses discussed in Sec.~\ref{sec:Tsallis_highenergy}. We display in Fig.~\ref{fig:betag} the behaviors of the beta function $\beta_{\bar{g}}$ as a function of $g$, and of the coupling $g$ as a function of~$\mu$. The results obtained here give a strong basis for the interpretation of previous experimental and phenomenological studies on QCD in terms of non-extensive statistics and thermofractals.

\section{Conclusions}
\label{sec:conclusions}

In this work we have introduced the non-extensive statistics in the form of Tsallis statistics of a quantum gas. Then, we have investigated the structure of a thermodynamical system presenting fractal properties showing that it naturally leads to Tsallis non-extensive statistics. Based on the scaling properties of gauge theories and thermofractal considerations, we have shown that renormalizable field theories lead to fractal structures, and these can be studied with Tsallis statistics. By using a recurrence formula that reflects the self-similar features of the fractal, we have computed the effective coupling and the corresponding beta function. The result turns out to be in excellent agreement with the one-loop beta function of QCD. Moreover, the entropic index, $q$, has been determined completely in terms of the fundamental parameters of the field theory. Finally, the result for $q$ is shown to be in good agreement with the value obtained by fitting Tsallis distributions to experimental data. These results give a solid basis from QCD to the use of non-extensive thermodynamics to study properties of strongly interacting systems and, in particular, to use thermofractal considerations to describe hadrons.

%%%%%%%%%%%%%%%%%%%%%%%%%%%
% Acknowledgments
\ack 
%%%%%%%%%%%%%%%%%%%%
A.D. and D.P.M. are partially supported by the Conselho Nacional de
Desenvolvimento Cient\'{\i}fico e Tecnol\'ogico (CNPq-Brazil) and by
Project INCT-FNA Proc. No. 464898/2014-5. A.D. is partially supported
by FAPESP under grant 2016/17612-7. The work of E.M. is supported by
the Spanish MINEICO under Grant FIS2017-85053-C2-1-P, by the FEDER
Andaluc\'{\i}a 2014-2020 Operational Programme under Grant
A-FQM-178-UGR18, by Junta de Andaluc\'{\i}a under Grant FQM-225, by
the Consejer\'{\i}a de Conocimiento, Investigaci\'on y Universidad of
the Junta de Andaluc\'{\i}a and European Regional Development Fund
(ERDF) under Grant SOMM17/6105/UGR, and by the Spanish Consolider
Ingenio 2010 Programme CPAN under Grant CSD2007-00042. The research of
E.M. is also supported by the Ram\'on y Cajal Program of the Spanish
MINEICO under Grant RYC-2016-20678.
%\vspace*{-1cm}
%\vfill \eject
%%%%%%%%%%%

%%%%%%%%%%%
\section*{References}
%\input{refs}
%\bibliographystyle{elsarticle-num}
%\bibliography{refs}
%%%%%%%%%%%

\end{document}